# Open Source Software in the Public Sector: 25 years and still in its infancy

**J. Linåker**
RISE Research Institutes of Sweden

**G. Robles**
Universidad Rey Juan Carlos

**D. Bryant**
Open Source Initiative

**S. Muto**
RISE Research Institutes of Sweden

*Abstract—* The proliferation of Open Source Software (OSS) adoption and collaboration has surged within industry, resulting in its ubiquitous presence in commercial offerings and shared digital infrastructure. However, in the public sector, both awareness and adoption of OSS is still in its infancy due to a number of obstacles including regulatory, cultural, and capacity-related challenges. This special issue is a call for action, highlighting the necessity for both research and practice to narrow the gap, selectively transfer and adapt existing knowledge, as well as generate new knowledge to enable the public sector to fully harness the potential benefits OSS has to offer.

■ **OPEN SOURCE SOFTWARE (OSS)** is today recognized as a pivotal building block in our common digital infrastructure [1]. A vast majority of today's software contains OSS, and the dependency on it in companies' codebases has grown considerably in recent years. Within public sector organizations, however, the adoption of OSS has thus far been rather limited in comparison, yet is on the rise [2].

A number of factors are driving public sector demand for OSS. Recent studies show that OSS adoption can bring a multitude of positive effects including economic growth, innovation, and competition. OSS has also been shown to bring benefits that are particularly salient in the public sector context, among them improved interoperability, transparency, and digital sovereignty [2-4].

In Europe, the importance of OSS for public sector transformation has recently been recognized both in joint Ministerial declarations [5] and by the European Commission through its establishment of an Open Source Program Office (OSPO), a function for promoting and enabling use and sharing of OSS [6]. The OSPO, introduced with inspiration from industry, is now spreading as an institutional construct and practice across several countries signaling an increased adoption of OSS. In Asia, notably in South Korea and China, OSS has already for some years featured strongly in industrial policy to support the competitiveness of the IT sector in those countries. In North America, the US government has long since established federal policies around the use and contribution to OSS. Today most notably in the wake of rising security and supply chain concerns for all forms of software, the current administration and federal agencies are discussing the role of OSS in security and seeking increased public-private partnership including discussions of reducing friction to enable broader participation in open source development and adoption. At the international level, interest in OSS can be noted among organizations such as the World Bank, World Health Organization, and the UN [7]. Initiatives such as the Phoenix and GovStack are also starting to emerge to provide an OSS infrastructure for the public sector, both in industrialized and developing countries.

The focus of this special issue is on how public sector organizations can adopt, develop, and collaborate on OSS, explore the conditions for success, and consider how software engineering practices may need to be adapted to the public context. The perspectives of a broad range of stakeholders were sought, including those of public sector organizations themselves, the industry, and wider OSS community, as each may have an vital role to play in this context.

OSS adoption in the public sector is an area that has thus far received relatively limited scholarly attention, despite a few notable exceptions [8-11]. This area poses specific challenges that differs from those encountered in the private sector context where software engineering practices have been more commonly investigated. In particular, the public sector domain necessitates the consideration of a broader range of factors, including a more complex set of motivations than those encountered in the private sector, technical capabilities, legally mandated constraints in the form of complicated procurement frameworks and practices, as well as limited or short-term policy incentives.

Below we present four different perspectives provided by the included papers of this special issue.

ENABLING MUNICIPAL COLLABORATION

Municipalities, typically characterized by their modest size, face resource and capacity constraints when seeking to develop and acquire technical solutions, including those utilizing OSS. Collaboration and resource pooling between municipalities, however, can facilitate the growth and shared capability to identify joint requirements, thus enabling the development of solutions that meet their collective needs.

In the study included in this special issue, Jullien and Viseur exemplify such pooling and collaboration through the case of CommunesPlone, an OSS project aimed at providing tools for public servants and modules for e-government solutions. The authors explore the journey of CommunesPlone from a grass-roots project initiated by two individuals from two municipalities to being hosted and developed through the publicly owned service company IMIO. The company is co-owned by 120 Wallonian municipalities, which it also serves, both in terms of enabling a collaborative requirements engineering process and offering necessary services to operationalize tools developed in the CommunesPlone OSS project.

Similar cases of public pooling and collaboration have also been studied through the use of non-profit foundations as commonly adopted in industry to enable co-opetition and neutral governance in multi-vendor OSS projects. The X-Roads OSS project provides one example where national governments join forces through NIIS, a co-owned foundation through which they acquire development from external suppliers [10]. OS2 (featured in the Requirements Engineering column of this special issue) is another where Danish municipalities collaborate on the requirements engineering, governance, and procurement of external development resources.

Yavuz et al. [12] describe this phenomenon through the lens of user-led foundations in the context of industry centered OSS-projects. Jullien and Viseur highlight that this type of collaboration moves away from the typical setting where the users are also the software developers in OSS projects. This emphasizes the need for further research and guidance for practice regarding how to collaborate on engineering and development of public sector-centered OSS solutions. Bridging requirements and software engineering practices with regulatory constraints, along with cultural and organizational aspects in the public sector, is of specific concern.

PROVISIONING OF NATIONAL SUPPORT

Public bodies' adoption strategies have long been debated in governmental, academic, and practitioner contexts. While governments have shown interest in the use and creation of OSS solutions –as can be seen from a large number of legal initiatives all around the globe–, the challenges faced are many. One first step is to promote the use of OSS by public administrations. The second step considers the development of OSS solutions for public administrations directly creating those solutions or by having them developed by means of procurement. A final step consists of the reuse of solutions from other public administrations.

Favario provides an experience report in this special issue on the Italian government's strategy to shift Public Administrations from a consumer-only role to a producer of OSS solutions, which is based on three key pillars: norms, tools, and community. The combination of these three elements is beginning to thrive in the Italian ecosystem, as evidenced by the number of administrations that started releasing their software with an OSS license. In addition, several other administrations have begun to reuse these solutions, as evidenced by the high ratio of software reuse of the software solutions in the national Italian OSS catalog.

All in all, the contribution by Favario shows that the transition to an OSS-aware public sector is a complex process that requires a careful strategy. Using an OSS solution is easy these days, but releasing a codebase introduces new challenges. Preparing for a release involves performing a whole series of operations that require advanced skills and careful planning. Once the code is released, it should be managed openly and transparently, and a clear strategy should be in place to prevent the situation from being unable to manage it.

It is also worth noting that even in those cases where a solution has been reused hundreds of times, it is still rare to see a real OSS community around it. This can be attributed to some circumstances, among others the novelty of the approach, which is not yet regulated in most public administrations and requires a change of mind to adopt the OSS approach. Further research from academia and novel approaches from practitioners are needed to address these aspects to find environments, technologies, governance structures, and legal frameworks that could make this possible.

The introduction of public sector OSPOs, or OSPO-like functions as exemplified by Favario, demonstrates how the public sector may draw lessons learned from the more mature industry context. Similar initiatives may be found (at the moment of writing) at national levels in, e.g., France, Germany, and Netherlands making or planning for similar efforts along those reported on in Italy. Still, this area has yet to mature, and although lessons may be drawn from the broader industry and community OSS ecosystems, practices may still have to be adapted to address and manage the challenges and conditions of the public sector context.

BUILDING INTERNAL CAPABILITIES BY LEVERAGING OPEN SOURCE COMMUNITIES

In the public sector, there is a long-standing practice of outsourcing and procuring technical resources, which can impede the development of internal competencies and affect organizational agility and absorptive capacity. This in turn, can have adverse effects on the organization's Digital Transformation (DX) journeys and their ability to reap the benefits of efficiency and innovation.

Within this special issue, Rudmark et al. present a case study focusing on Entur, a public company that falls under the Norwegian Ministry of Transport and Communications, with a specific focus on offering sales and ticket solutions for the railways and travel planner for the public transport of Norway. The study details how Entur was able to develop its DX capabilities by reducing its dependence on externally procured tools and resources, and instead, harnessing the multiplier effect and innovation opportunities provided by OSS and Open Data communities.

Entur had previously procured an external proprietary travel planning system, but stalemate responses from the vendor on change requests, along with substantially underestimated development costs and limited quality on releases, convinced Entur to look elsewhere, leading to the adoption of the OpenTripPlanner, an OSS equivalent. Although the OSS solution was not immediately fit for purpose, Entur was able to implement national open data standards, improve algorithms, and integrate with internal systems by dedicating internal developers to working with the community and building trust and influence on the OSS project's development and requirements engineering efforts.

Entur deems that leveraging the crowd provides access to knowledge and innovation that would otherwise be beyond their reach. Yet, the study highlights that investing in OSS may not be an inexpensive undertaking in the short term as it may require significant investment to realize its potential benefits. Hence, the authors recommend that OSS should primarily be considered as a sourcing option in situations where the organization has ambitious goals for the solutions and the willingness to undertake the necessary organizational and cultural changes.

The cost-benefit analysis, as highlighted, is a problematic assessment that requires thorough knowledge from a technical, user, and regulatory perspective. The consideration of OSS as an option in acquisition processes needs to be improved due to limited internal capabilities. Here, efforts such as those reported by Favario in this special issue may help, e.g., by provisioning procurement guidelines, software catalogs, and assessment support. Future research is needed to shed more light on how OSS can be considered and acquired, as the DX journey for the public sector has only just commenced.

ENHANCING TRANSPARENCY THROUGH SOFTWARE BILL OF MATERIALS

Cybersecurity has become one of the most urgent challenges in all sectors of society in recent years. Countries worldwide grapple with developing effective public policies and implementing best practices to support improved security for all concerned. Beginning in 2021, governments around the world, notably the European Union, the United States, Japan, and Canada, have either expressed their intent or have codified a requirement that all software developed for their governments –and, in some cases, critical industries-- employ improved supply chain practices including Supply Bills of Material (SBOM) to enhance and deepen transparency into some obfuscated software.

In his study in this special issue, Holbrook makes a case for increased transparency and improved security in software development in the US federal government environment. The study draws on field experience through a legal and intellectual property lens while working with government practitioners. The paper provides a contextual understanding of SBOM and its relationship to software licenses and the practical value of transparency, includes recommendations for technical policies, and grounds the reader in existing authorities and citations on the matter for the US. Today those US policies address federal agency acquisition but are anticipated to cascade to regulated industries in the future.

The paper proposes a triad of benefits in using SBOM; enhancing cybersecurity, avoiding the misuse of OSS, and complying with US law. It also suggests that adopting development best practices ensure OSS code remains high-quality as well as secure and compliant. The ability to respond quickly to vulnerabilities, with the Log4j event as an example, is cited as the high value of having an auditable train of components when a significant cyber incident occurs in OSS.

The use of SBOMs is an emerging best practice. Standards development for Supply Chain and SBOM are an open discussion at this writing. Although the study and extensive citations in the study are centered around the US experience, the practices and insight shared should have value to readers globally. The practical benefits of adding SBOMs into the software development process align with technical and security best practices, both domestically and internationally.

SUMMARY AND CALL TO ACTION

We know that the adoption of OSS in the Public Sector has been on the rise for some time. This is seen both in the use of the development methodology to create new capabilities specific to public sector operations as well as the adoption and use of existing OSS applications and platforms. Most recently, the COVID-19 pandemic brought with it an explosion of software development in the public interest as OSS, harnessing the power of collaborative development to innovate quickly.

We have observed a marked increase in all corners of the public sector; in schools and universities, local and provincial governments, and national governments. In many countries, OSS has become an essential part of their Information and Communications Technology (ICT) strategy by more efficiently providing service to citizens. In some cases, OSS is embedded in strategies for economic development. We know OSS is solving real problems in public health, public safety, humanitarian relief, and other government-related services.

Given this recent momentum, it would seem an opportune time to document and share the experience of OSS projects within the public sector, including in-depth analysis of the challenges and the resulting benefits to both organizations and the citizens they serve. However, the editors note that the number of submissions for this issue, particularly in terms of academic contributions, has been lower than anticipated. The majority of submissions came from practitioners, which is also reflected in the final selection. Practitioner use cases are particularly valuable because it reflects lived experience underpinning the practical telling of the use case. Sharing these use cases can also serve to coalesce the kind of communities around these projects that we know today are key to their success. However, there is also a need for more formal research and case studies to complement these practitioner narratives, which can enable others to learn from these experiences in a more structured manner. This field of study presents numerous research opportunities, especially in relation to public institutions. Overcoming the challenges associated with the implementation of OSS solutions in the public sector has the potential to yield significant benefits, similar to what has already been observed in industry.

As noted in the contributions of this special issue, the situation of OSS in the public sector lies well behind the private sector. Despite numerous efforts and initiatives that have been undertaken all over the world, some dating more than 25 years, public administrations still struggle to achieve fundamental aspects, such as building a community around OSS solutions. In comparison, groups of volunteers (the canonical examples are the Apache project and the Linux kernel in its early years) already formed healthy communities 30 years ago. More recently, the private sector has also shaped thriving communities around OSS projects, first in hybrid forms (e.g., the more recent forms of the Linux kernel development) and more recently as a conglomeration of companies (as in OpenStack or the automotive COVESA alliance). We thus argue that OSS in the public sector is still in its infancy and that there is a significant road ahead of bringing it to levels that the OSS community and the private sector have already conquered.

Overall, there exists a substantial body of experience that can serve as a valuable starting point for future research in this field. We were grateful for the opportunity to gain the many insights provided by reviewing the papers for the special edition, and it is our hope and encouragement that it may inspire further research. There is much more to be known about the numerous potential benefits of open technology adoption, not the least of which is an acceleration of innovation in the public sector - something which has thus far mostly benefited the private sector.

## ACKNOWLEDGMENT

We are incredibly grateful for the support we received from several individuals who made this special issue possible and successful, including the anonymous reviewers, authors, and IEEE Software's editorial staff.

**Dr. Johan Linåker** is a Senior Researcher at RISE Research Institutes of Sweden. His research interests include empirical software engineering research in industry and public sector in the context of open innovation. He is specifically interested in the areas of Open Source Software and Open Data and how these can serve as a source of open innovation and help create value for software-intensive organizations, independently and through an ecosystem context. Johan also holds an interest in requirements engineering and product management. Contact him at johan.linaker@ri.se.

**Dr. Gregorio Robles.** is a full professor at the Universidad Rey Juan Carlos, Madrid, Spain. His research focuses on free/open source software. Robles received his Ph.D. in empirical software engineering research on free/open source software in 2006. He co-authored the 2002 FLOSS study while working at the University of Maastricht, The Netherlands. He was the general chair of the IEEE International Conference on Software Maintenance and Evolution in 2018 and of the IEEE International Conference on Mining Software Repositories in 2021. He was also the program co-chair of the 2016 and 2017 Open Source Software Conferences, among others. He is a Member of the IEEE. Contact him at grex@gsyc.urjc.es

**Deborah Bryant,** is Director of Public Policy for the non-profit Open Source Initiative. She has held interest in the successful adoption of open source in the public sector including its use in regional ICT strategies throughout her career spanning the public, private and university sectors since 2005. She specializes in OSS related public policy including cybersecurity and has an interest in community health and sustainability. Her published research includes "Study of Open Source for Cyber Security in the Energy Sector" (US Department of Energy, 2011); "Case Study & Model Development Project, International Public Administration Open Source Code Communities", with Price Waterhouse Coopers of Spain for CENATIC Foundation (2010) Contact her at deb.bryant@opensource.org

**Sachiko Muto** is a Senior Researcher at RISE Research Institutes of Sweden, and the Chair of OpenForum Europe (OFE). She originally joined OFE in 2007 and served for several years as Director with responsibility for government relations and then as CEO. Sachiko is a co-author of "The impact of Open Source Software and Hardware on technological independence, competitiveness and innovation in the EU economy", a study published by the European Commission in 2021. She has degrees in Political Science from the University of Toronto and the London School of Economics and has been a guest researcher at UC Berkeley and TU Delft. Contact her at sachiko.muto@ri.se.